\documentclass[runningheads]{llncs}

\usepackage{subfigure}
\usepackage{comment}
\usepackage{soul}
\usepackage{graphicx}
\usepackage{pgfplots}
\usepackage{graphicx}

\usepackage{lipsum}
\usepackage{booktabs}

\usepackage{subcaption}
\usepackage{tikz}
\usepackage{array}
\usepackage{amsmath}

\RequirePackage{amsfonts}
\RequirePackage{mathtools}
\usepackage{acronym}
\usepackage{xcolor}
\usepackage{hyperref}
\hypersetup{
  colorlinks=true,
  linkcolor=black,
  urlcolor=black,
  citecolor=black,
}
\usepackage{balance}

\usepackage[numbers]{natbib}

\usepackage{glossaries}
\newacronym{ai}{AI}{Artificial Intelligence}
\newacronym{ris}{RIS}{Reconfigurable Intelligent Surface}
\newacronym{b5g}{B5G}{Beyond-Fifth-Generation}
\newacronym{ula}{ULA}{Uniform Linear Array}
\newacronym{mdp}{MDP}{Markov Decision Process}
\newacronym{aod}{AoD}{Angle-of-Departure}
\newacronym{aoa}{AoA}{Angle-of-Arrival}
\newacronym{ml}{ML}{Machine Learning}
\newacronym{mlp}{MLP}{Multi-Layer Perceptron}
\newacronym{rl}{RL}{Reinforcement Learning}
\newacronym{drl}{DRL}{Deep Reinforcement Learning}
\newacronym{dqn}{DQN}{Deep Q-Networks}
\newacronym{ddpg}{DDPG}{Deep Deterministic Policy Gradient}
\newacronym{td3}{TD3}{Twin Delayed Deep Deterministic Policy Gradient}
\newacronym{ppo}{PPO}{Proximal Policy Optimization}
\newacronym{sac}{SAC}{Soft-Actor Critic}
\newacronym{sinr}{SINR}{Signal-to-Noise-plus-Interference Ratio}
\newacronym{los}{LoS}{Line-of-Sight}
\newacronym{nlos}{NLoS}{Non Line-of-Sight}
\newacronym{iid}{i.i.d.}{independent and identical distribution}
\newacronym{qos}{QoS}{Quality-of-Service}
\newacronym{qsos}{QSoS}{Quality \& Security of Service}
\newacronym{fqos}{FQoS}{Fairness \& Quality of Service}
\newacronym{fqoss}{FQoSS}{Fairness \& Quality of Service Score}
\newacronym{bs}{BS}{Base Station}
\newacronym{mu}{MU}{Mobile User}
\newacronym{ue}{UE}{User Equipment}
\newacronym{ssr}{SSR}{Sum Secrecy Rate}
\newacronym{sr}{SR}{Secrecy Rate}
\newacronym{jfi}{JFI}{Jain Fairness Index}

\newcommand{\etal}{\textit{et al}. }
\newcommand{\ie}{\textit{i.e.}, }

\newif\ifcomments

\usepackage{color}
\usepackage[symbol]{footmisc}

\begin{document}

\title{A Fairness-Aware Strategy for B5G Physical-layer Security Leveraging Reconfigurable Intelligent Surfaces}

\author{A. Pierron\inst{1}, M. Barbeau\inst{2}, L. De Cicco\inst{3}, J. Rubio-Hernan\inst{1}, J. Garcia-Alfaro\inst{1}}

\institute{SAMOVAR, T\'el\'ecom SudParis, Institut Polytechnique de Paris, Palaiseau, France
\and
Carleton University, School of Computer Science, Ottawa, Canada
\and
Politecnico di Bari, Dipartimento di Ingegneria Elettrica e dell'Informazione, Italy
}
  
\maketitle

\begin{abstract}

Reconfigurable Intelligent Surfaces are composed of physical elements that can dynamically alter electromagnetic wave properties to enhance beamforming and lead to improvements in areas with low coverage properties.
When combined with Reinforcement Learning techniques, they have the potential to enhance both system behavior and physical-layer security hardening. In addition to 
security improvements, it is crucial to consider the concept of \textit{fair communication}. Reconfigurable Intelligent Surfaces must ensure that User Equipment units receive their signals with adequate strength, without other units being deprived of service due to insufficient power.  In this paper, we address such a problem. We explore the fairness properties of previous work and propose a novel method that aims at obtaining both an efficient and fair duplex Reconfigurable Intelligent Surface-Reinforcement Learning system for multiple legitimate User Equipment units without reducing the level of achieved physical-layer security hardening. In terms of contributions, we uncover a fairness imbalance of a previous physical-layer security hardening solution, validate our findings and report experimental work via simulation results. We also provide an alternative reward strategy to solve the uncovered problems and release both code and datasets to foster further research in the topics of this paper.\\

\end{abstract}

\section{Introduction}

    \noindent The transition toward \gls*{b5g} and 6G networks introduces challenges with respect to performance, security, and scalability~\cite{mao2023security,wang2023road,chen20235g}. As wireless infrastructures become increasingly heterogeneous and densely deployed, achieving reliable communication with stringent energy-efficiency requirements has become a central concern. Within this context, \glspl*{ris} are a recently proposed technology that allows reconfiguring the propagation environment through real-time control of electromagnetic waves~\cite{di2020smart,dai2020reconfigurable,liu2021reconfigurable,ETSI_GR_RIS_001_2025_main_picture}.
    
    Unlike conventional architectures, \gls*{ris} comprise arrays of program\-mable sub-wavelength elements whose reflection properties can be dynamically adjusted. This capability enables sophisticated beamforming and channel manipulation strategies that can significantly enhance network capacity and resilience~\cite{basar2019wireless, pan2022overview}. Potential applications span a broad range, including coverage extension in complex urban scenarios~\cite{di2020smart}, energy-efficiency improvements via optimized propagation paths~\cite{huang2019reconfigurable}, and reinforcement of physical-layer security mechanisms~\cite{do2022physical}.
    
    The latter aspect is particularly critical. Traditional cryptographic methods operating at higher protocol layers are insufficient against vulnerabilities inherent to the physical layer. For instance, unprotected key-exchange procedures remain susceptible to replay or injection attacks, which may subsequently undermine higher-layer protections. Physical-layer security, enabled by \gls*{ris}, offers a compelling approach to mitigate such threats by shaping the wireless environment itself to secure initial exchanges~\cite{do2022physical}.
    
    Despite their advantages, the integration of \glspl*{ris} into practical systems entails substantial challenges. Effective deployment requires taking into account unit mobility, channel variability, and real-time adaptability. These constraints are exacerbated by the limited computational and energy resources available to \gls*{ris}, which are typically realized as low-power embedded devices. Moreover, in multi-user or multi-unit scenarios, fairness refers to the equitable allocation of resources, performance gains, or quality of service across participating entities. Ensuring such fairness while maintaining overall system performance introduces a complex, multi-objective optimization problem.
    
    To address these challenges, we advocate the use of \gls*{drl}. Unlike traditional model-driven optimization methods, \gls*{drl} is well suited for highly dynamic, non-linear environments, as it enables the autonomous learning of control policies through direct interaction with the system~\cite{arulkumaran2017deep, wang2022deep, tang2025deep}. In particular, our approach incorporates fairness-aware mechanisms to ensure that unit-specific Quality of Service (QoS) is preserved without compromising aggregate performance or security.
    
    The key contributions of this work are summarized as follows:  (i) we uncover a fairness limitation in related work using RL-assisted physical-layer \glspl*{ris} security hardening, (ii) we propose a new reward strategy that tackles the problem, (iii) we validate our claims, both formally and practically; and (iv) we conclude that the new strategy can provide an equivalent level of security hardening without renouncing fairness properties.
    
    The remaining sections are organized as follows. 
    Section~\ref{sec:relatedWork} reviews related literature. Section~\ref{sec:ProblemStatement} formulates the problem and outlines modeling assumptions. Section~\ref{sec:rewards} details the proposed methodology. Section~\ref{sec:experiments} presents the evaluation results. Section~\ref{sec:futureWork} provides some perspectives and future directions for research. Section~\ref{sec:conclusion} concludes the paper and provides future work perspectives.

\section{Background and Related Work}
\label{sec:relatedWork}

    \subsection{Reconfigurable Intelligent Surfaces}
    \label{subsec:ris}

        Reconfigurable Intelligent Surface (\gls*{ris}) constitutes a novel class of engineered planar structures designed to manipulate electromagnetic (EM) waves in a nearly-passive manner~\cite{bjornson2024introduction, di2020smart, basar2019wireless}. Unlike conventional active antenna arrays, which radiate signals by means of dedicated radio-frequency (RF) chains, a \gls*{ris} is composed of a large number of low-cost, sub-wavelength scattering elements, often referred to as meta-atoms or unit cells, that can locally control the phase (and, in some implementations, the amplitude) of incident EM waves. Through external control signals, these elements impose programmable reflection coefficients, thereby reshaping the propagation environment.
        
        We consider a \gls*{ris} with $N$ reflecting elements. The baseband equivalent of the reflection matrix can be written as
        \begin{equation}
            \boldsymbol{\Theta} = \text{diag}\left(\xi_1 e^{j\theta_1},\xi_2 e^{j\theta_2},\ldots, \xi_{N} e^{j\theta_N}\right),
            \label{eq:theta_matrix}
        \end{equation}
        where $\theta_n \in [0, 2\pi)$ denotes the programmable phase shift of the $n$-th element. Assuming negligible reflection loss, each coefficient has unit modulus, i.e., \newline $|\xi_{n}| =1 ~\space \forall n \in \{1,\dots, N\}$. By jointly tuning these parameters, the \gls*{ris} can reinforce constructive interference at an intended receiver or enforce destructive interference toward unintended ones.
        
        The fundamental mechanism enabling these gains is the coherent superposition of multiple reflected signals. When the phase shifts $\{\theta_n\}$ are optimized to align with the direct and scattered channels, the received signal components add constructively, boosting the effective channel gain. Conversely, when $\{\theta_n\}$ are chosen to introduce specific phase mismatches, the scattered signals can cancel out undesired components, thus reducing interference levels or even creating nulls in certain spatial directions~\cite{wu2019intelligent, bjornson2020power}. This dual ability makes \glspl*{ris} attractive for interference-limited scenarios, such as dense urban deployments and cell-edge communication.
        
        From a physical standpoint, the array gain offered by a \gls*{ris} scales approximately quadratically with $N$ under favorable propagation conditions, provided that phase coherence is maintained across all elements~\cite{bjornson2024introduction}. This property allows large surfaces to partially compensate for the so-called ``double path-loss'' effect caused by the cascaded transmitter–\gls*{ris}–receiver channel. At the same time, the destructive interference capability provides a tool for improving spatial reuse, mitigating inter-cell interference, and enhancing physical-layer security~\cite{najafi2020physics, renzo2020smart}.
        
        The technology underpinning \gls*{ris} draws heavily from metamaterials and frequency-selective surfaces, which have long been studied in applied electromagnetics~\cite{liu2021reconfigurable}. In contrast to fully active relays, which require power-hungry RF chains, \glspl*{ris} typically consume power only for the control circuitry, making them an energy-efficient solution for future wireless networks.

    \subsection{Deep Reinforcement Learning}
        \label{subsec:DRL}
        The dynamic and complex nature of wireless networks with \gls*{ris} requires advanced control strategies capable of real-time adaptation. In this context, \gls*{drl} is a particularly suitable methodology, extending classical reinforcement learning methods to handle high-dimensional state and action spaces typical of such environments.
        
        Reinforcement Learning (RL) is rooted in the principles of optimal control and dynamic programming. Early work by Bellman on dynamic programming \cite{bellman1957dynamic} and Kalman on optimal control \cite{kalman1960contributions} laid the foundation for modern RL. In RL, an agent interacts with an environment, typically formalized as a \gls*{mdp} \cite{puterman2014markov}, to maximize the cumulative reward. At each time step $t$, the agent observes the system state $s_t$, takes an action $a_t$, and receives a reward $r_t$ as the system transitions to a new state $s_{t+1}$. The agent's objective is to maximize the expected discounted return, expressed through the state-action value function:
        \[
        Q^\pi(s,a) = \mathbb{E}\left[\sum_{t=0}^\infty \gamma^t r_t \,\bigg|\, s_0 = s, a_0 = a, a_t \sim \pi(s_t)\right],
        \]
        where $\pi$ is the policy mapping states to actions, and $\gamma$ is the discount factor. 
        
        RL methods can be categorized along two key dimensions: \emph{on-policy} vs \emph{off-policy} and \emph{value-based} vs \emph{policy-based}. On-policy methods learn from data generated by the current policy, while off-policy methods can learn from data generated by different policies. Value-based methods focus on estimating the value function, from which a policy is derived, while policy-based methods directly optimize the policy.
        
        The advent of deep neural networks~\cite{rumelhart1986learning, goodfellow2016deep} gave rise to \gls*{drl}, enabling RL to scale to complex, high-dimensional problems. \gls*{dqn} is an off-policy, value-based method that uses a deep neural network to approximate the Q-function, achieving human-level performance in discrete action spaces~\cite{DQ-learning}.
        
        For continuous action spaces, such as those encountered in \gls*{ris} control, \gls*{ddpg} \cite{ddpg} is particularly relevant. \gls*{ddpg} is an off-policy, policy-based method that combines deterministic policy gradients with an actor-critic framework. It uses a deep neural network for the actor (policy) and another for the critic (value function), with Experience Replay to improve sample efficiency~\cite{adam2011experience}. This makes \gls*{ddpg} well-suited for the continuous control requirements of \gls*{ris}-assisted communication systems.
    
        \begin{figure*}[!t]
                    \centering
                    \includegraphics[width=.9\linewidth]{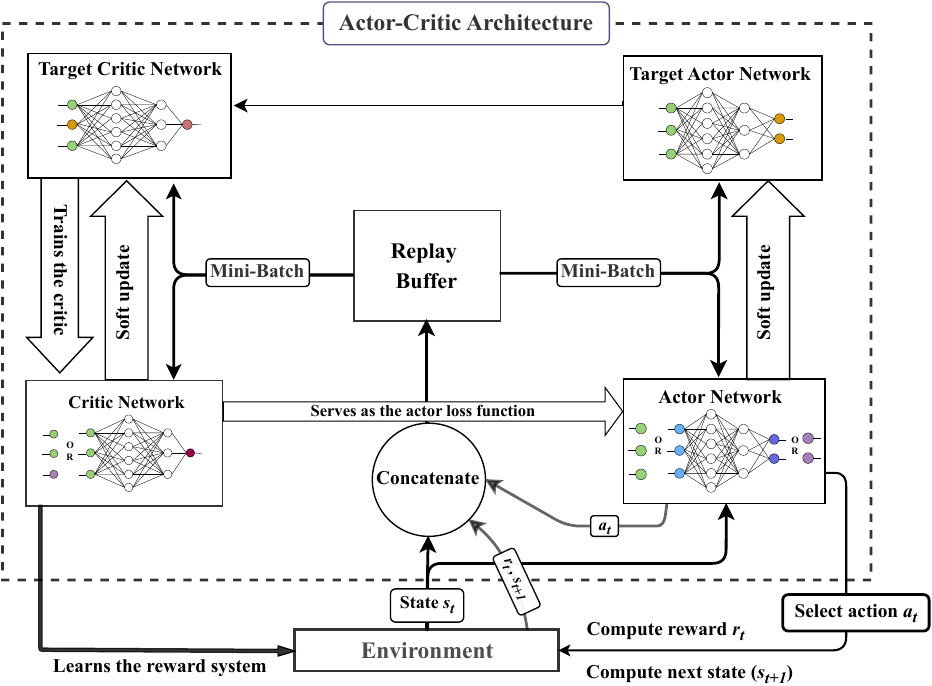}
                    \caption{General Actor–Critic architecture employed in \gls*{ddpg}.}
                    \label{fig:network_architecture}
            \end{figure*}

    \subsection{Related Work}
    \label{subsec:related_work}

        The intersection of \gls*{drl} and \gls*{ris}-assisted communication has gained considerable research attention in recent years. Liu et al.~\cite{liu2021reconfigurable} provide a broad survey of the field, highlighting the adaptability of \gls*{drl}-based methods to dynamic environments and their ability to optimize multiple performance objectives simultaneously.
        
        Several works demonstrate the effectiveness of \gls*{drl} in enhancing system performance. Nguyen et al.~\cite{nguyen2022achievable} showed that \gls*{drl}-optimized \gls*{ris} configurations can achieve notable gains in energy efficiency when compared to traditional non-learning, model-based schemes. Huang et al.~\cite{huang2020reconfigurable} propose a \gls*{drl}-driven framework for maximizing data rates while mitigating interference, while Peng et al.~\cite{Peng2022} extend this approach to duplex communication, addressing both uplink and downlink transmissions under eavesdropping threats. Similarly, Yang et al.~\cite{yang2020deep} incorporate quality-of-service constraints into \gls*{drl}-based optimization, underscoring the flexibility of the approach.
        
        Despite these advances, several critical gaps remain. First, most existing studies rely on simplified channel models that do not fully capture the propagation characteristics of large-scale \gls*{ris}~\cite{bjornson2024introduction}. Second, fairness in multi-unit scenarios has been largely overlooked, with optimization efforts focused primarily on aggregate system metrics. Finally, systematic comparisons across different \gls*{drl} algorithms for \gls*{ris} optimization remain limited, leaving open questions regarding algorithm suitability for distinct communication objectives.

    \begin{table}[!t]
        \caption{Acronyms and Notation} 
        \label{tab:notation}
        \centering
        \begin{tabular}{|c|l|}
            \hline
            \textbf{Acronym} & \textbf{Description} \\
            \hline
            BS & Base Station \\
            DRL & Deep Reinforcement Learning \\
            UE & User Equipment  \\
            SINR & Signal-to-Noise-plus-Interference Ratio \\
            RIS & Reconfigurable Intelligent Surface \\
            \hline
            \textbf{Notation} &  \\
            \hline
            $N$ & Total number of elements in the \gls*{ris} \\
            $k$ & Number of \gls*{ue} units \\
            $l$ & Number of eavesdroppers \\
            $D_{i}$ & Downlink information rate for \gls*{ue}$_i$ \\
            $U_{i}$ & Uplink information rate of \gls*{ue}$_i$ \\
            $R_{i}$ & Secrecy rate for unit $i$ \\
            $E_{i,j}$ & Information rate of eavesdropper $j$ over unit $i$ \\
            $C$ & Sum secrecy rate \\
            $R_{QoS}$ & Quality-of-Service Reward \\
            $R_{fm}$ & Fairness Minimax Reward \\
            $R_{fs}$ & Fairness Smoothed Reward \\
            \hline
        \end{tabular}
    \end{table}

\section{System Considered and Problem Statement}
\label{sec:ProblemStatement}    

    In this section, we define the considered system and clarify the problem we want to study. Table~\ref{tab:notation} summarizes the main acronyms and notations used in our work. We consider a scenario where a \gls*{bs} with $N_t$ transmit antennas and $N_r$ receiving antennas communicates with $k$ legitimate \gls*{ue} units, each modeled as a single-antenna transceiver. A direct \gls*{los} link between the \gls*{bs} and the \glspl*{ue} is assumed to be blocked by an obstacle. To overcome this limitation, a \gls*{ris} composed of $N$ reflecting elements is deployed to assist both downlink and uplink communications. For tractability, we assume the network topology lies in a planar configuration \cite{huang2019reconfigurable, huang2020reconfigurable, Peng2022}.

    \subsection{Physical-layer Model}
    \label{sub:physical_layer}
        
\begin{figure}[!t]
            \centering
            \includegraphics[width=\linewidth]{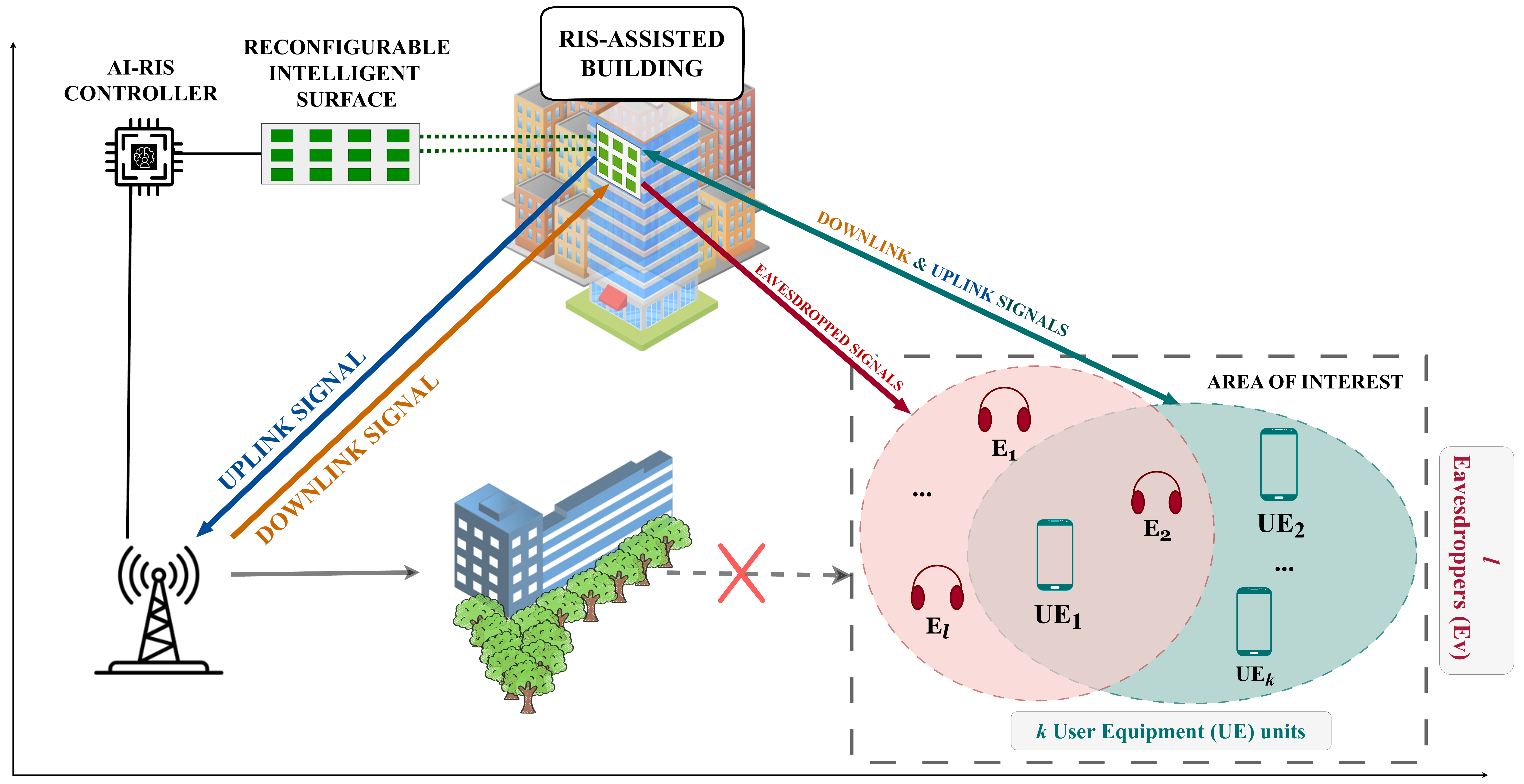}
            \caption{Considered \gls*{ris}-\gls*{drl} environment with multiple \glspl*{ue}  and eavesdroppers.}
            \label{fig:env_multiple_user}
        \end{figure}

        We build upon the general \gls*{ris}-\gls*{drl} setup proposed by Peng \etal~\cite{Peng2022} (cf. Figure~\ref{fig:env_multiple_user}). The system includes both legitimate \gls*{ue} and potential eavesdroppers, all of which are fixed in location throughout the duration of an episode. We retain the propagation model, interference treatment, and \gls*{sinr} computation used in \cite{Peng2022}, while adapting the channel modeling to better capture the physical behavior of \gls*{ris}, as detailed in~\cite{bjornson2024introduction}.  
        
            Both the \gls*{bs} and \gls*{ris} are modeled as Uniform Linear Arrays (ULAs). The \gls*{los} channel components follow the physically consistent formulation in~\cite{bjornson2024introduction}, and we do not consider the \gls*{nlos} components of the signal for simplification.

            We compute the array response channel $\boldsymbol{H_{t}^r}$ between a transmitter $\boldsymbol{t}$ and receiver $\boldsymbol{r}$ as
            \begin{equation}
            \boldsymbol{H_{t}^r}  = 
            \begin{bmatrix}
            1 \\
            \vdots \\
            e^{-j2\pi \tfrac{(A_r-1)\Delta \sin\varphi_r}{\lambda}}
            \end{bmatrix}
            \begin{bmatrix}
            1 & \cdots & e^{-j2\pi \tfrac{(A_t-1)\Delta \sin\varphi_t}{\lambda}}
            \end{bmatrix},   
            \end{equation}
            where $\lambda$ is the wavelength, $\varphi_t$ and $\varphi_r$ denote the \gls*{aod} and \gls*{aoa}, respectively, $\Delta$ is the antenna spacing, and $A_t$ and $A_r$ are the number of antennas at transmitter and receiver.
            
            The large-scale channel gain $\beta$ for the \gls*{bs}–\gls*{ris}–\gls*{ue} cascade is derived from~\cite{bjornson2024introduction}:
            \begin{align}
                \beta_t &= \frac{ G_t(\phi_t,\theta_t) A_m}{4 \pi d_t^2}, \qquad
                \beta_r = \frac{ G_r(\phi_r,\theta_r) A_m}{4 \pi d_r^2}, \\
                \beta   &= N^2 \beta_t \beta_r 
                         = N^2 \frac{ G_t(\phi_t,\theta_t) G_r(\phi_r,\theta_r) A_m^2}{(4 \pi d_t d_r)^2},
            \end{align}
            where $d_t$ and $d_r$ denote the distances from transmitter and receiver to the \gls*{ris}, respectively, $A_m$ is the area of each reflecting element (approximated as $(\lambda/4)^2$), and $G_t,G_r$ are respectively the antenna gain functions at the transmitter and the receiver, with $\phi_t,\phi_r$ azimuth angles and $\theta_t,\theta_r$ elevation angles.
            As shown in~\cite{bjornson2024introduction}, the \gls*{ris} gain scales quadratically with $N$, making very large surfaces suitable to offset the cascaded path losses. Throughout this work, we assume the \gls*{ris} is a \gls*{ula} constituted of isotropic antennas with a normalized gain, $G_t=G_r=1$.
            This assumption simplifies our analysis by focusing on the inherent properties of the \gls{ris} and use cases of mobile communications such as \gls*{b5g} networks~\cite{Balanis2016, bjornson2024introduction}.

    \subsection{Threat Model}
    \label{subsec:threat_model}
        
    We consider eavesdroppers as the primary adversaries in the system, representing the main physical-layer threat to \gls*{ris}-assisted communication. Eavesdroppers aim to intercept downlink or uplink transmissions whenever they fall within the high-gain region of the beam. Following~\citet{Peng2022,yang2020deep}, we assume $l$ eavesdroppers attempting to capture the transmitted signals. Although this model is simplified, it effectively captures a realistic security concern, as \gls*{ris} beamforming can inadvertently create unintended high-gain lobes toward external nodes. These lobes can expose sensitive transmissions to interception by unauthorized receivers. Therefore, secrecy analysis is essential to ensure robust and confidential communication in \gls*{ris}-assisted systems.
    
    \subsection{Controller Model}
        \label{subsec:ControllerModel}
            
            Our \gls*{drl} agent controls both the \gls*{bs} beamforming matrix $\boldsymbol{W} \in \mathbb{C}^{N_t \times k}$ and the \gls*{ris} diagonal phase-shift matrix $\boldsymbol{\Theta} \in \mathbb{C}^{N \times N}$, whose entries are unit-modulus complex exponentials. The transmit power constraint at the BS is enforced by projecting $\boldsymbol{W}$ as
            
            \begin{equation}
            \Pi(\boldsymbol{W}) = \begin{cases}
                        \boldsymbol{W}, & \text{if } \mathrm{Tr}(\boldsymbol{W}\boldsymbol{W}^H) \leq P_{\text{max}}, \\
                        \frac{\sqrt{P_{\text{max}}}}{\|\boldsymbol{W}\|_F}\boldsymbol{W}, & \text{otherwise},
                    \end{cases}
            \end{equation}
            where $\|\cdot\|_F$ denotes the Frobenius norm.  
            
            We employ the \gls*{ddpg} algorithm~\cite{ddpg} for continuous control (cf Section \ref{subsec:DRL}). The observation vector $s_t$ at time $t$ includes the cascaded channel state, phase noise, previous slot rates, and the action applied at $t-1$, following~\citet{Peng2022}.
            
    \subsection{Problem Formulation}
        
        We can now formalize the optimization problem. First, we introduce common terms and the baseline reward function.
        
            For each unit $i$, the achievable downlink and uplink data rates are
            \begin{align}
                D_{i} &= \log_2 \!\left(1 + \text{SINR}_{i}^{\text{D}}\right), \label{eq:R_B}\\
                U_{i} &= \log_2 \!\left(1 + \text{SINR}_{i}^{\text{U}}\right), \label{eq:R_S}
            \end{align}
            where $\text{SINR}_{i}^{\text{D}}$ and $\text{SINR}_{i}^{\text{U}}$ are the downlink and uplink SINRs, respectively. The data rate overheard by eavesdropper $j$ from unit $i$ is
            \begin{equation}
                E_{i,j} = \log_2\!\big(1 + \text{SINR}^E_{d,i,j}\big) + \log_2\!\big(1 + \text{SINR}^E_{u,i,j}\big),
            \end{equation}
            where $\text{SINR}^E_{d,i,j}$ and $\text{SINR}^E_{u,i,j}$ denote the downlink and uplink SINR components as seen by the eavesdropper. These quantities form the building blocks of the secrecy-aware reward functions.
            
            Following~\citet{Peng2022}, the secrecy rate for unit $i$ is
            \begin{equation}
                R_{i} = \left[ D_{i} + U_{i} - \max_{j} E_{i,j} \right]^+, \label{eq:R_i}
            \end{equation}
            where $[x]^+=\max(0,x)$. The baseline reward, i.e., the \gls*{ssr}, is then defined as
            \begin{equation}
                C = \sum_{i=1}^{k} R_{i}. \label{eq:SSR}
            \end{equation}
            This reward emphasizes overall capacity and secrecy, without fairness or QoS guarantees, and is widely adopted in the literature~\cite{huang2020reconfigurable,bjornson2024introduction}. In our experiments, we denote this reward as the \emph{baseline}.
            \\
            
            The overall objective is to maximize a chosen reward function under the system constraints. For the baseline \gls*{ssr} reward, this yields:
            \begin{equation}
            \begin{array}{cl}
            \underset{\mathbf{W}, \boldsymbol{\Theta}}{\text{maximize}} & C(\mathbf{W}, \boldsymbol{\Theta}, \boldsymbol{\Phi}, \mathcal{H}) \\
            \text{subject to} & \mathrm{Tr}(\mathbf{W}\mathbf{W}^H) \leq P_{\text{max}}, \\
                              & 0 \leq \theta_m < 2\pi, \quad m = 1,\dots,N.
            \end{array}
            \end{equation}
            where $\mathcal{H}$ denotes the set of all channels involved in the communication scenario. Alternative reward formulations incorporating fairness (cf. Section~\ref{subsec:fairness_reward}) or QoS guarantees can replace $C$, enabling flexible control of system objectives.

    \subsection{Fairness Model}
    \label{subsec:fairness_theory}
        
        In addition to secrecy and capacity, fairness among units is a key performance dimension. Without fairness, strong channels or favorable spatial locations may monopolize resources, while disadvantaged units experience poor service. To quantify this, we adopt the \gls*{jfi}~\cite{jain1984quantitative}:
        \begin{equation}
        \mathrm{I}_{\text{Jain}} = \frac{\left(\sum_{i=1}^{k} R_i\right)^2}{k \cdot \sum_{i=1}^{k} R_i^2},
        \label{eq::jain_fairness}
        \end{equation}
        where $R_i$ is the secrecy rate of unit $i$. The index ranges from $1/k$ to 1, with unity indicating perfect equal distribution of the rewards between each participant, meaning perfect fairness.
        
        A fundamental trade-off, named \textit{price of fairness}, exists between maximizing total capacity and ensuring fairness \cite{bertsimas2011price}. As highlighted in~\cite{bjornson2024introduction}, optimizing purely for sum rate often leads to  uneven allocations, since resources are concentrated on units with favorable channels. Conversely, enforcing strict fairness (e.g., maximizing the minimum secrecy rate) generally lowers the aggregate capacity. 
        
        This trade-off is of practical importance in \gls*{b5g}-6G scenarios where both high throughput and equitable unit experience are essential. For example, ultra-reliable low-latency communication services may prioritize fairness, while enhanced mobile broadband services lean toward capacity maximization. Our study explicitly integrates this trade-off into the reward functions explored in Section~\ref{sec:rewards}, thereby enabling control over system-wide fairness levels.

\section{Reward Functions and Fairness Improvements}
\label{sec:rewards}

    The objective of this section is to address the fairness issues discussed in Section~\ref{subsec:related_work}. We also design a new reward strategy that maximizes the legitimate data rate transmitted to \gls*{ue}, while maintaining the same level of security enhancement. We seek to ensure that each unit receives a minimum usable data rate. The goal is to provide a fair service among all units despite the non-linearity of our optimization problem. Next, we explore further the reward functions selected for our work.

    \subsection{Quality-of-Service-Aware Reward Function}
    \label{subsec::yang_reward}
    The reward function in~\citet{yang2020deep} introduces unit-specific \gls*{qos} constraints. The reward, denoted here as the \textit{QoS} reward, is given by
    \begin{align}
        R_{QoS} &= \sum_{i=1}^{k} (R_{i} - \mu_1  p_i^S\ - \mu_2p_i^U) , \label{eq:QoS}
    \end{align}
    where $\mu_1$ and $\mu_2$ are positive constants balancing penalties for matching respectively the security and transmission criteria, and
    \begin{align*}
                    p_i^S &=
                    \begin{cases}
                    1, & \text{if } R_i < \epsilon_i^{S}, \\
                    0, & \text{otherwise}
                    \end{cases}\\
                    p_i^U &=
                    \begin{cases}
                    1, & \text{if } D_i +U_i < \epsilon_i^{U}, \\
                    0, & \text{otherwise}
                    \end{cases}
    \end{align*}
    with $\epsilon_i^S, \epsilon_i^U \in \mathbb{R}^+$ denoting respectively the target secrecy rate for the $i$-th \gls*{ue} and the target data rate.  
    
    This formulation introduces a first notion of service by penalizing configurations where individual units fail to meet their minimum requirements. However, once thresholds are satisfied, the system remains free to allocate resources disproportionately, favoring specific units to maximize overall reward. This limitation motivates the need for fairness-aware extensions.
    
    \subsection{Fairness-aware Reward Function}
    \label{subsec:fairness_reward}

    As discussed in Section~\ref{subsec:fairness_theory}, balancing overall system capacity and security with fairness among units represents a critical tradeoff in \gls*{ris}-assisted communication. The baseline reward function provides no explicit incentive to distribute resources equitably, while the \textit{QoS} reward in~\eqref{eq:QoS} ensures only a minimum rate per unit without addressing fairness beyond threshold satisfaction. To overcome these limitations, we introduce fairness-aware reward formulations that explicitly account for equitable resource distribution in both downlink and uplink.
    
    We define the following intermediate terms for unit $i$:
    
    \begin{align*}
    R_{di} &= \left[ D_i - \max_{\substack{j \in [1:l]}} \log_2\left(1 + \operatorname{SINR}^E_{d,i,j}\right) \right]^+, \\
    R_{ui} &= \left[ U_i - \max_{\substack{j \in [1:l]}} \log_2\left(1 + \operatorname{SINR}^E_{u,i,j}\right)\right]^+, \\
    R_d &= \min_{\substack{i \in [1:k]}} R_{di}, \\
    R_u &= \min_{\substack{i \in [1:k]}} R_{ui}, \\
    \end{align*}
    
    By construction, $R_d$ and $R_u$ represent the bottleneck unit rates in each direction, thereby embedding fairness into the optimization process. Based on these terms, we propose two candidate reward functions:
    
    \begin{align}
    R_{fm} &= k \cdot (R_d + R_u), \label{eq:Rfm} \\
    R_{fs} &= \left(\frac{\sum_{i=1}^{k} R_{di}}{1 + \operatorname{std}(R_{di})}
    \right)^{p_f}+ \left(\frac{\sum_{i=1}^{k} R_{ui}}{1 + \operatorname{std}(R_{ui})}\right)^{p_f}, \label{eq:Rfs}
    \end{align}  
    where $R_{fm}$ emphasizes the minimum guaranteed rate across units by using a minimax approach \cite{bjornson2024introduction}, $R_{fs}$ penalizes large rate disparities by incorporating the standard deviation of per-unit performance and $p_f \in \mathbb{N}$ is a hyperparameter. 
    
    Equation~\eqref{eq:Rfm} implements a hard minimax objective by depending only on the bottleneck rates $\min_i R_{di}$ and $\min_i R_{ui}$. Although this gives strict worst-case guarantees, the nested $\max$, $\min$ and the $[\cdot]^+$ operator produce a highly non-smooth, piecewise reward surface. In RIS-assisted beamforming, small changes in phase shifts or power allocations can abruptly change which unit is worst-off, causing discontinuous jumps in $R_{fm}$. Such discontinuities and the underlying non-convexity significantly impair gradient-based or sample-limited RL algorithms, as they increase gradient variance, induce slow or unstable convergence, and make the policy highly sensitive to initialization. Moreover, a pure minimax objective is inherently conservative: raising the minimum often forces large sacrifices in aggregate throughput. 

    By contrast, Equation~\eqref{eq:Rfs} acts as a smooth surrogate that explicitly trades off aggregate throughput and dispersion. The numerator rewards overall secrecy performance while the denominator penalizes unequal allocations; the additive constant prevents numerical instability when the spread is small. This ratio (i) yields a denser, more informative learning signal because every unit contributes to the reward (unlike the minimum which depends only on the single worst unit), (ii) produces a smoother optimization landscape that reduces gradient variance and improves training stability in the high-dimensional, nonconvex RIS action space, and (iii) allows the designer to tune the fairness/throughput tradeoff via the exponent $p_f$. For these reasons $R_{fs}$ is a practically relevant, trainable compromise when using learning-based controllers, whereas $R_{fm}$ remains appropriate if one requires strict, provable guarantees on the worst-case unit.
    
    We name these two reward functions respectively the $fm$ and $fs$ rewards for \textit{fairness minimax} and \textit{fairness smoothed}.

\section{Experimental Framework and Results}
\label{sec:experiments}

    In this section, we present the experimental framework and simulation results. We first describe the design of the framework and the adopted algorithms, and then analyze the system performance in terms of capacity, fairness, and robustness against eavesdroppers. To ensure reproducibility, the full code is publicly released in a companion repository.\footnote{\url{https://github.com/alex-pierron/CTRL_RIS}} 
    
    \subsection{Framework Features}
    
    The proposed framework provides a \gls*{drl} environment tailored for duplex communications between a \gls*{bs} and multiple \gls*{ue} units, assisted by a \gls*{ris}. The environment is implemented in Python, with all \gls*{drl} algorithms developed using PyTorch. The framework is openly available to reproduce our experiments and to support further research on the application of \gls*{drl} to \gls*{ris}-aided communication systems. 
    
    The system includes monitoring tools to evaluate learning dynamics and performance metrics, thereby providing a complete overview of the training process. In this study, we adopt the \gls*{ddpg} algorithm~\cite{ddpg}, following the actor–critic architecture previously used in~\cite{Peng2022,huang2020reconfigurable}. Both the actor and the critic use a \gls*{mlp} \cite{hornik1989multilayer}. The state space is unchanged from Peng \etal~\cite{Peng2022}, while the algorithm itself follows its original implementation~\cite{ddpg,fujimoto2018addressing}, except for modifications to the reward functions in order to reflect the objectives defined in Section~\ref{sec:rewards}.
    
    \subsection{Experimental Setup}
    
    We consider three representative scenarios: one without eavesdroppers and two with eavesdroppers. The environment remains consistent in all scenarios. The \gls*{bs}, positioned at coordinates $(0,0)$, communicates with the \gls*{ris} located at $(20,100)$ in both directions. Each scenario is trained over $40\,000$ steps. The \gls*{bs} is equipped with $N_t = N_r = 4$ antennas, while the \gls*{ris} consists of $N = 36$ elements. We consider two legitimate \gls*{ue} units, \ie $k=2$. In the scenarios with eavesdroppers, we assume $l$ is also set to two eavesdroppers. We use a wavelength $\lambda = 0.1$\,m, a maximum emitting power of $20$\,dBm for the \gls*{bs} and $100$\,mW for the legitimate units.
    
    Scenario~1 corresponds to a baseline case without eavesdroppers, used to illustrate the trade-off between capacity and fairness in the absence of external threats. Scenario~2 introduces eavesdroppers positioned close to the legitimate users: the \gls*{ue} units are placed at $(147,20)$ and $(169,80)$, while the eavesdroppers are placed at $(140,40)$ and $(150,75)$. Scenario~3 considers a different spatial arrangement, where the legitimate units are at $(124,40)$ and $(138.5,81)$ and the eavesdroppers at $(132,50)$ and $(128,85)$.  
    
    Each scenario is trained with a dedicated controller, since \gls*{drl} methods are highly sensitive to hyperparameter tuning. The training and neural network main settings are summarized in Table~\ref{tab:nn_hyperparam} and also include the main hyperparameters for the reward functions. The full detailed hyperparameter value list is also available at our public GitHub repository.

    \begin{table}[!t]
    \centering
    \caption{Main hyperparameters used}
    \label{tab:nn_hyperparam}
    \begin{tabular}{l c}
    \toprule
    \textbf{Parameter} & \textbf{Value} \\
    \midrule
    Actor learning rate ($\delta_a$) & $1 \times 10^{-5}$ \\
    Critic learning rate ($\delta_c$) & $1 \times 10^{-3}$ \\
    Soft update coefficient ($\tau$) & $5 \times 10^{-4}$ \\
    Replay buffer size & $5000$ \\
    Batch size & $96$ \\
    Actor update frequency & Every two steps \\
    Critic update frequency & Every step \\
    Training steps & $40\,000$ \\
    Target Data Rate $\epsilon_i^U$ & 2.5 \\
    Target Secrecy Rate $\epsilon_i^S$ & 1.5 \\
    $\mu_1$ & 2 \\
    $\mu_2$ & 2 \\
    $p_f$ & 2 \\
    \bottomrule
    \end{tabular}
    \end{table}


\begin{figure}[!hbt]
 \centering
 \subfigure[Scenario 1]{
    \includegraphics[width=0.47\textwidth]{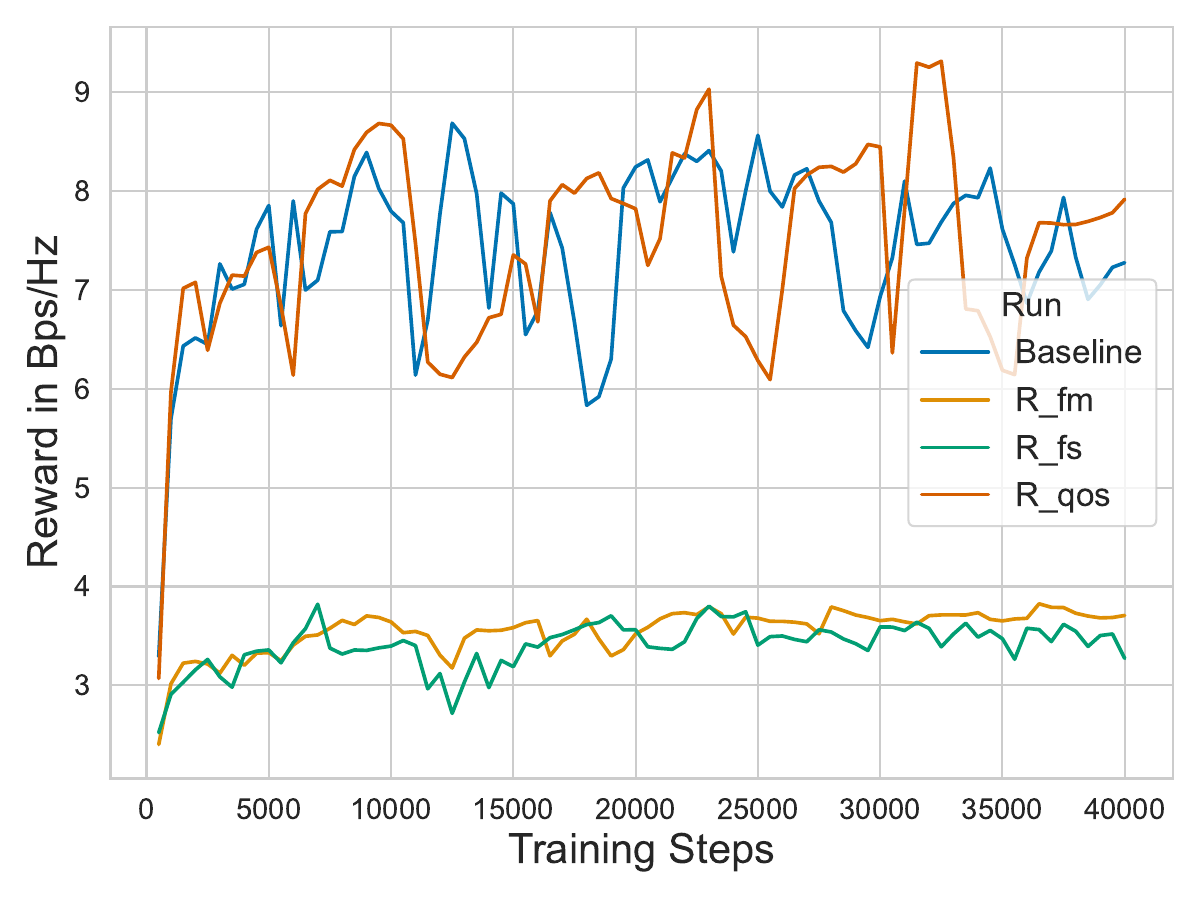}
 } 
 \subfigure[Scenario 2]{
    \includegraphics[width=0.47\textwidth]{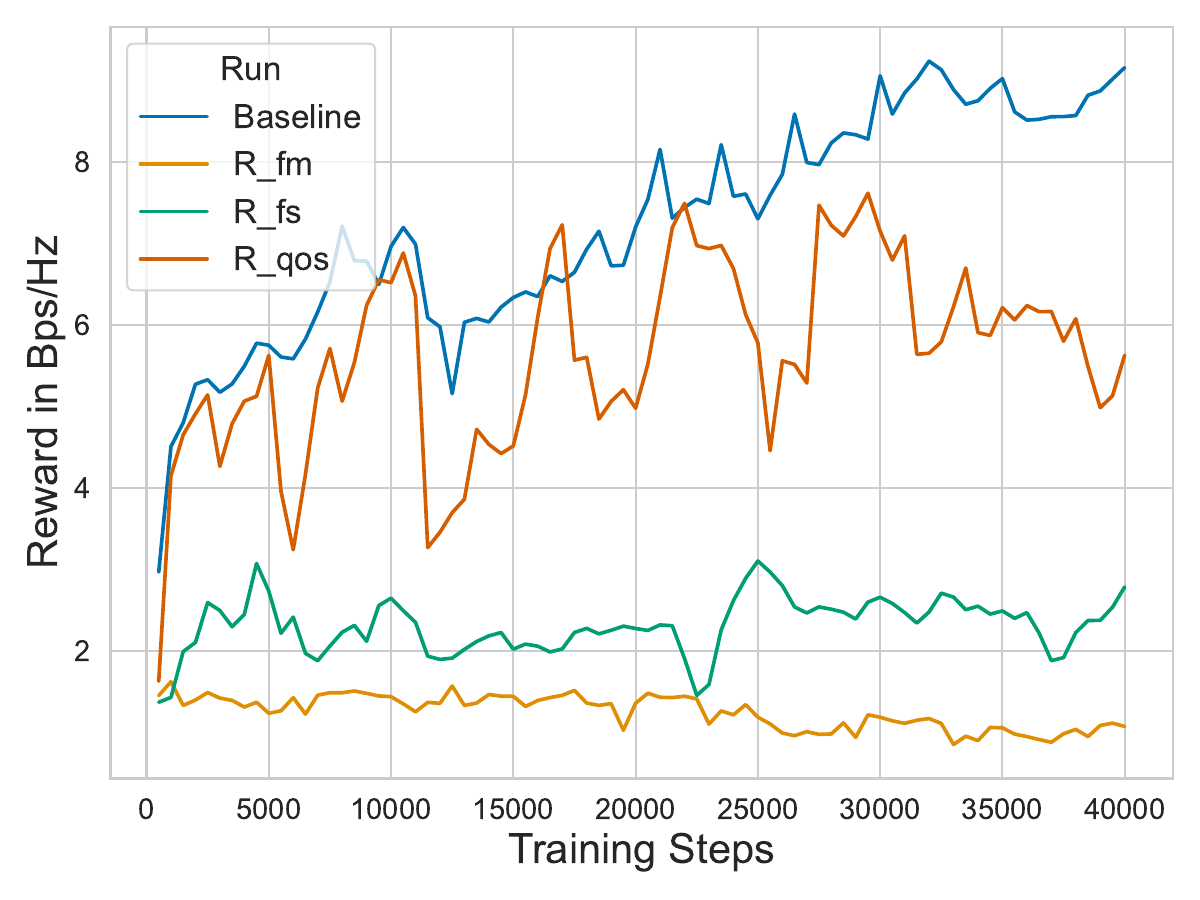}
 }
 \subfigure[Scenario 3]{
    \includegraphics[width=0.47\textwidth]{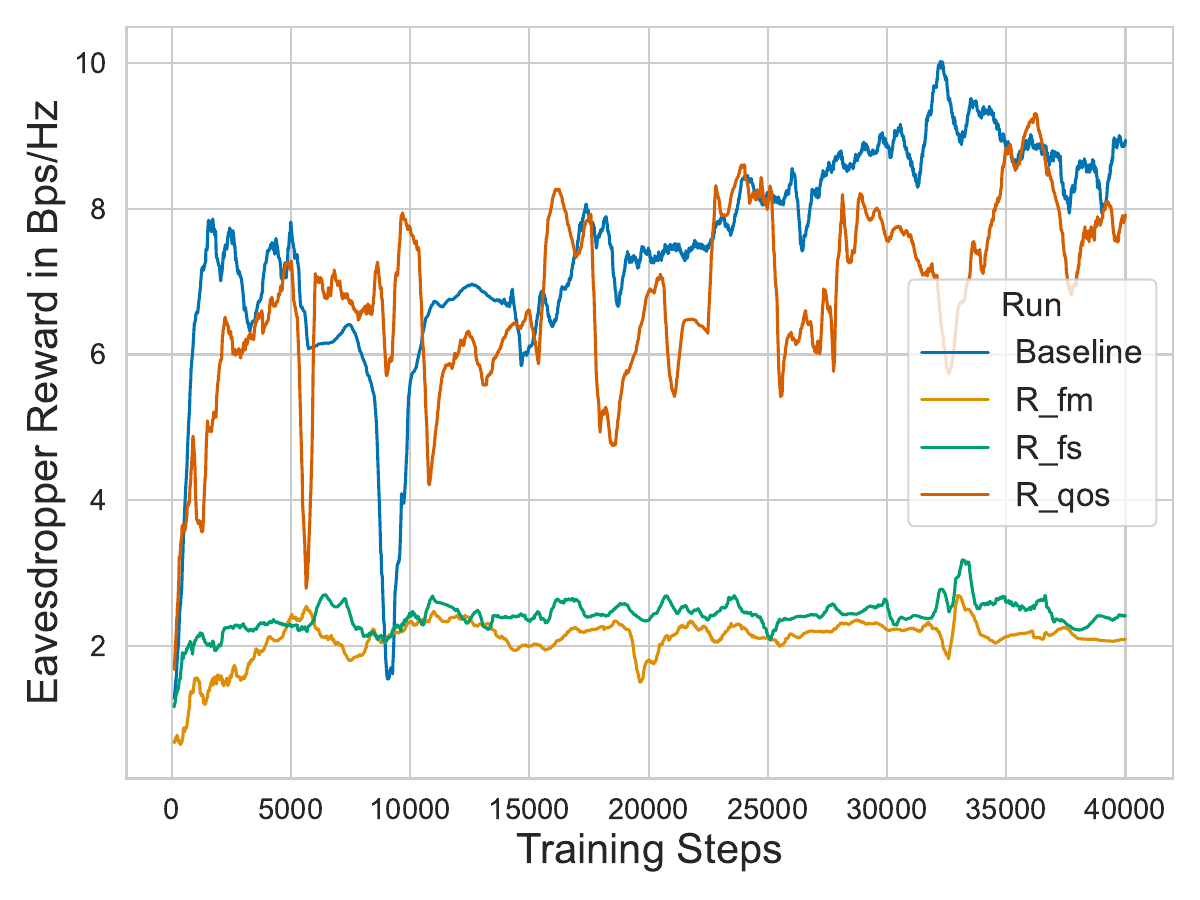}
 }
 \caption{Evolution of system capacity with respect to the \textit{baseline} reward in Scenarios 1--3. Each plot represents the baseline reward over time (local average) in Bps/Hz. Results are smoothed with a rolling window of size $500$.}
\label{fig::results_capacity}
\end{figure}

\begin{figure}[!hbt]
 \centering
 \subfigure[Scenario 1]{
    \includegraphics[width=0.47\textwidth]{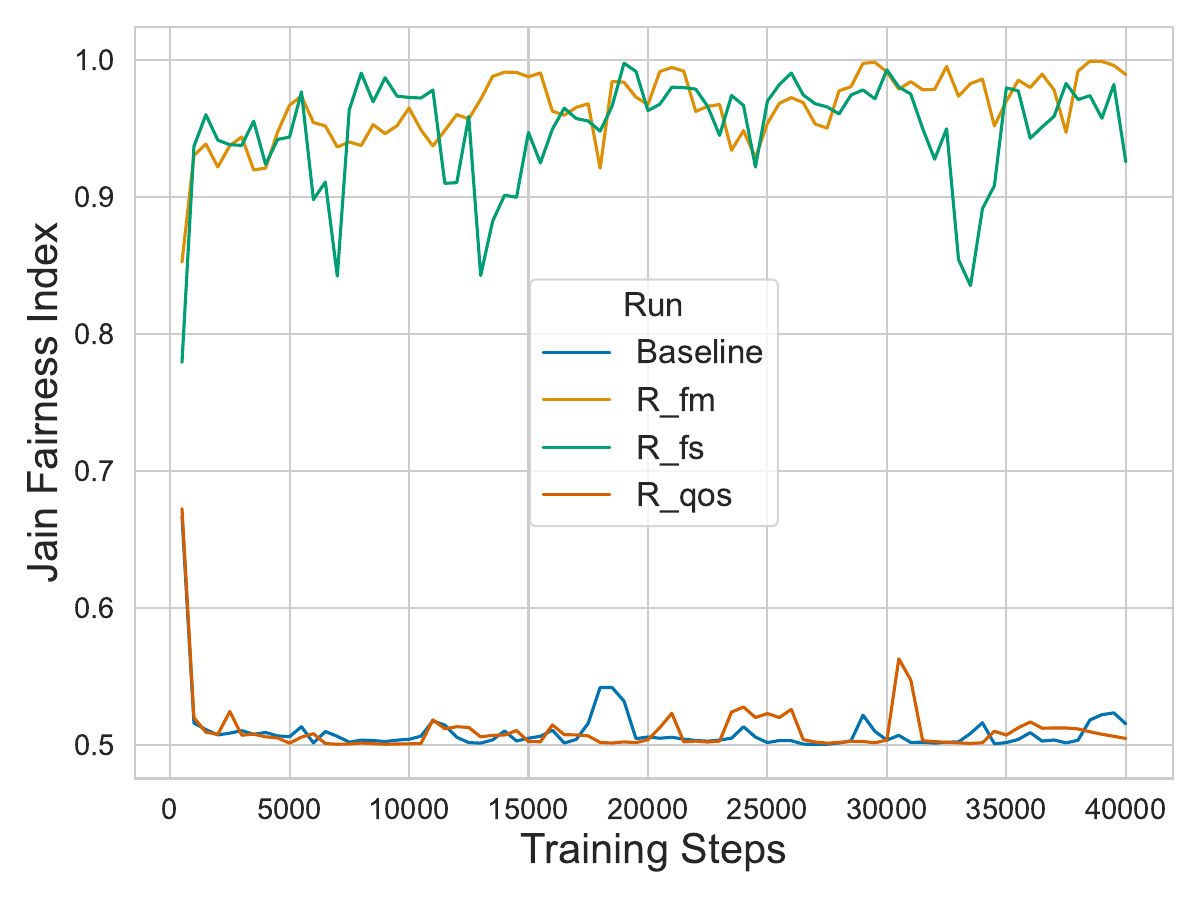}
 } 
 \subfigure[Scenario 2]{
    \includegraphics[width=0.47\textwidth]{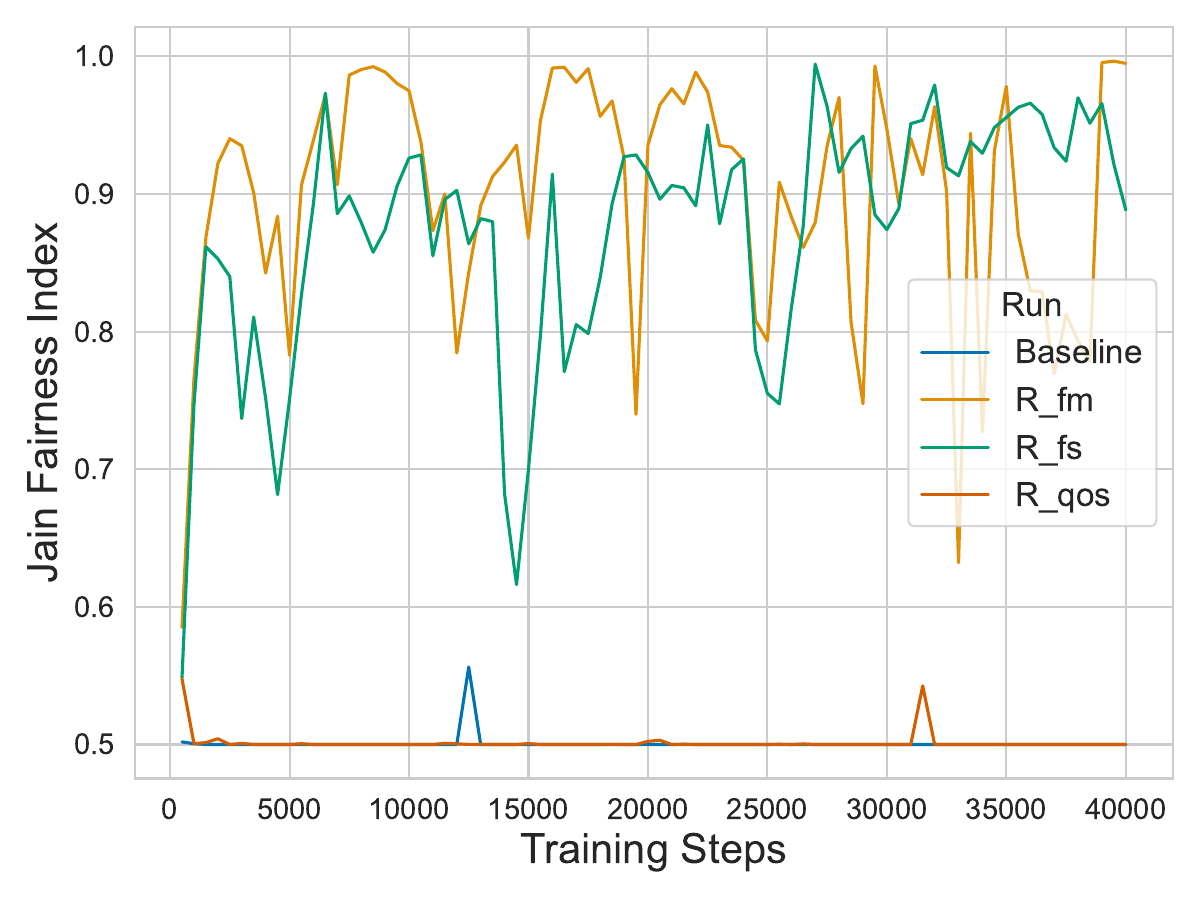}
 }
 \subfigure[Scenario 3]{
    \includegraphics[width=0.47\textwidth]{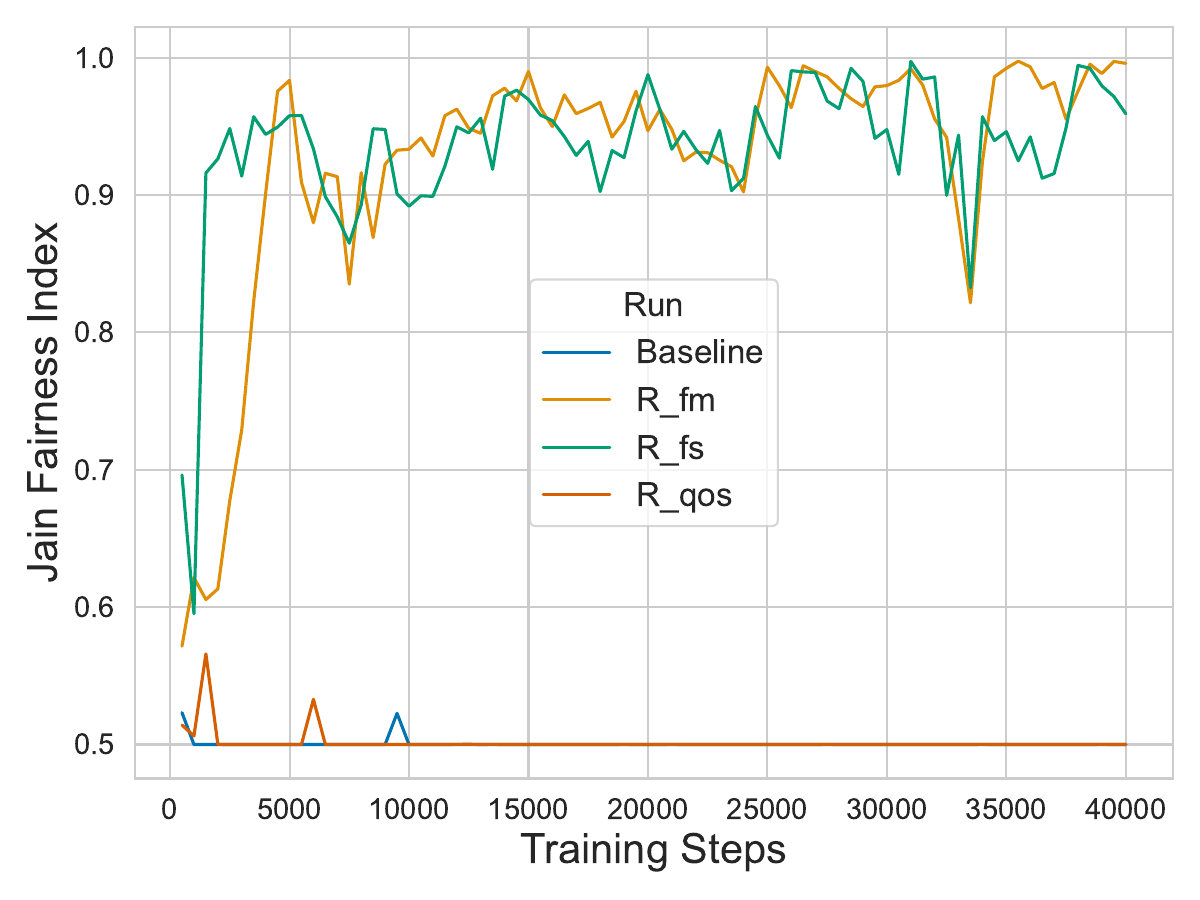}
 }
 \caption{Evolution of Jain Fairness Index (\gls*{jfi}) in Scenarios 1--3. Each plot represents user JFI over time (local average). Results are smoothed with a rolling window of size $500$.}
\label{fig::results_fairness}
\end{figure}

\begin{figure}[!hbt]
 \centering
 \subfigure[Scenario 2]{
    \includegraphics[width=0.47\textwidth]{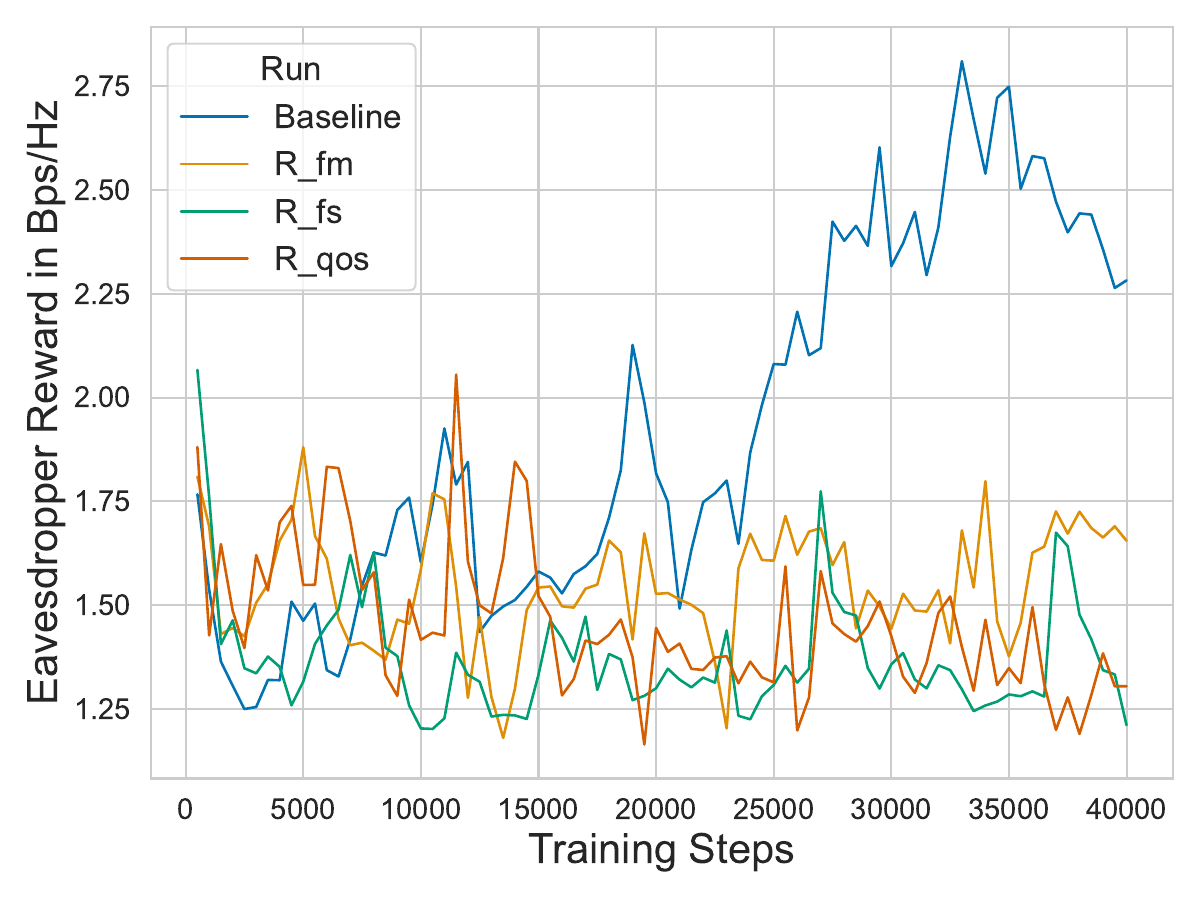}
 } 
 \subfigure[Scenario 3]{
    \includegraphics[width=0.47\textwidth]{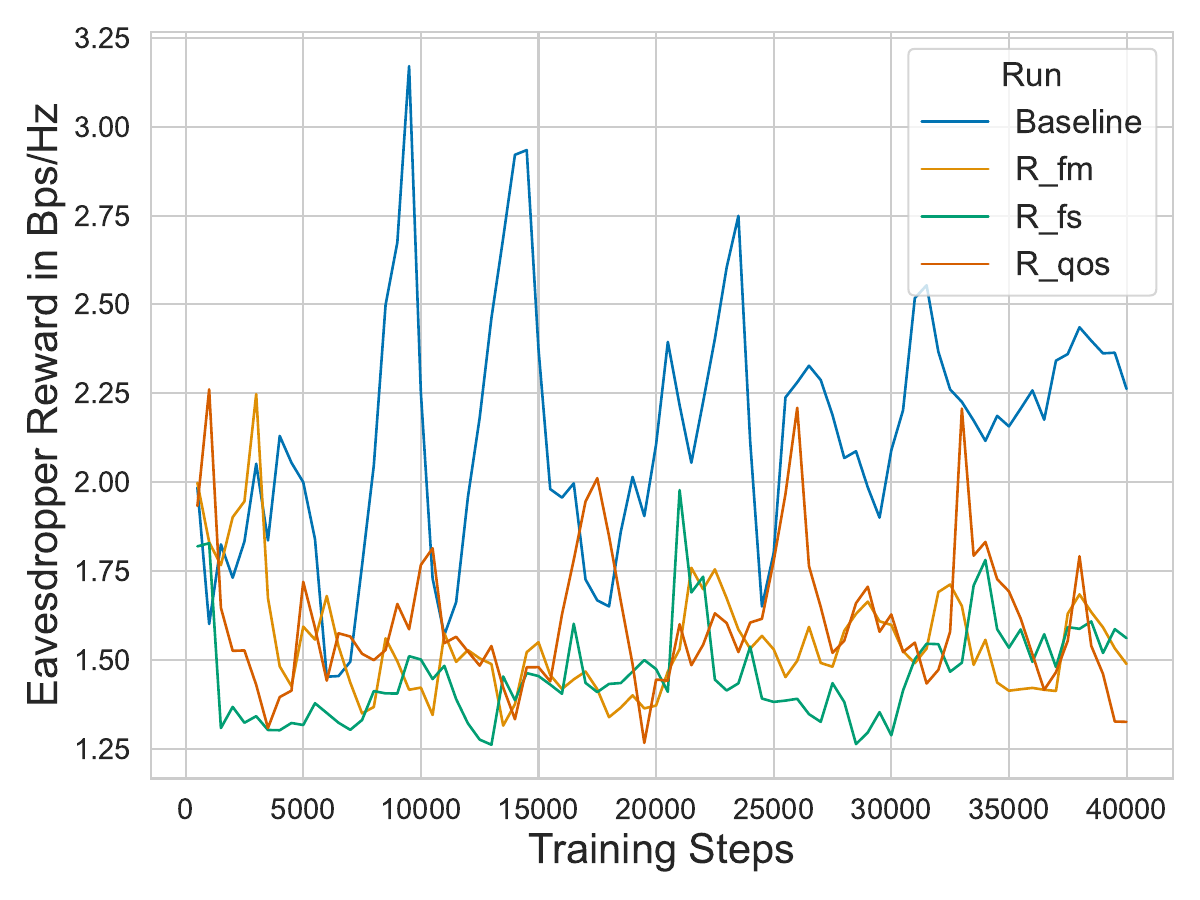}
 }
 \caption{Eavesdroppers' rewards with respect to the \textit{baseline} reward in Scenarios 2 and 3. Each plot represents the eavesdropper reward over time (local average) in Bps/Hz. Results are smoothed with a rolling window of size $500$.}
\label{fig::results_eaves}
\end{figure}

    \subsection{Results}

    The performance of the different reward functions introduced in Section~\ref{sec:rewards} is evaluated in terms of capacity, fairness, and robustness against eavesdroppers. All metrics are normalized with respect to the \textit{baseline} reward function to enable consistent comparisons.
    
    The evolution of system capacity is shown in Figure~\ref{fig::results_capacity}, while Figure~\ref{fig::results_fairness} reports the fairness measured using the Jain Fairness Index (JFI). In Scenario~1, the \textit{baseline} and \textit{QoS} rewards achieve higher capacity, almost twice as high on average as that obtained with the $fm$ and $fs$ rewards. However, this gain comes at the expense of fairness: the average JFI is around $0.5$, which corresponds to a completely unfair allocation. By contrast, the $fm$ and $fs$ rewards lead to substantially higher fairness, with an average JFI close to $0.9$, though at the cost of reduced capacity. 
    Notice that this result is expected and confirms the generality of the \textit{price of fairness} property described in Section~\ref{subsec:fairness_theory}, \ie the price to pay to make the allocation fair~\cite{bertsimas2011price}.
    
    The same trade-off is observed in Scenarios~2 and 3, which involve eavesdroppers. While reaching high fairness becomes more difficult, the $fm$ and $fs$ functions still achieve a significantly better balance compared to the baseline and \gls*{qos} functions, confirming the robustness of their fairness-oriented design.
    
    The analysis of eavesdropper performance, shown in Figure~\ref{fig::results_eaves}, provides further insights. When using the \textit{baseline} reward, the eavesdroppers benefit from improved performance in parallel with the legitimate users. Conversely, with $fm$ and $fs$, the rewards obtained by the eavesdroppers remain consistently lower, indicating that these reward functions also provide improved security by limiting information leakage.

    The $fm$ and $fs$ reward functions enhance fairness and robustness against eavesdroppers, but at the cost of raw system capacity. This trade-off highlights the importance of carefully designing reward mechanisms to meet system-level objectives. The experimental evaluation demonstrates that fairness-aware reward formulations can effectively balance the competing demands of system capacity, equitable resource allocation, and physical-layer security, providing valuable insights for the design of practical \gls*{ris}-assisted communication systems.

    Additional experimental data, detailed plots, and TensorBoard logs are available in the companion repository\footnote{\url{https://github.com/alex-pierron/CTRL_RIS}} to facilitate further analysis and reproducibility.

\section{Future Directions for Research}
\label{sec:futureWork}

    Several promising research directions emerge from this work. First, the development of general controllers capable of operating in diverse environment configurations represents a significant challenge. Current \gls*{drl} methods require careful hyperparameter tuning and are highly sensitive in order to properly generalize. Training a single controller that can adapt to varying network topologies, user distributions, and threat models would substantially improve the practical applicability of \gls*{drl}-based \gls*{ris} control. This generalization problem could be addressed through curriculum learning \cite{bengio2009curriculum, narvekar2020curriculum}, imitation learning \cite{zare2024survey} or transfer learning techniques \cite{pan2009survey, zhu2023transfer} that leverage knowledge from previously trained controllers. Since \gls*{drl} controllers would ultimately be embedded on the \gls*{ris}, their inference performance should be systematically compared with non-learning, model-based optimization methods to assess practical deployment trade-offs.

    Second, a systematic comparison with non-learning, model-based optimization baselines would provide valuable insights into the trade-offs between \gls*{drl} approaches and traditional optimization methods. While \gls*{drl} offers adaptability to dynamic environments, model-based techniques such as semidefinite programming or alternating optimization may provide more interpretable solutions and computational guarantees. Understanding when each approach is preferable would help guide practical system design.

    Third, the \gls*{ris} model could be refined to better capture practical hardware constraints. Decoupling uplink and downlink elements would enable more realistic duplex communication modeling \cite{liu2021reconfigurable, wu2024intelligent}, while incorporating amplitude control and accounting for quantization errors would provide more faithful representations of real-world \gls*{ris} implementations.

    Finally, dynamic scenarios using relevant mobility patterns for the users and eavesdroppers present an important extension. Investigating how fairness-aware reward functions perform under such dynamic conditions while maintaining fairness guarantees would be a relevant research direction.

\section{Conclusion}
\label{sec:conclusion}

We explored the concept of \textit{fair communications} in \gls*{drl}-enabled \gls*{ris} environments. We assumed that \glspl*{ris} must ensure that \gls*{ue} units receive their signals with adequate strength, without other \gls*{ue} being deprived of service due to insufficient power. We investigated this problem and examined the fairness properties of previous work. We proposed a novel method that aims at obtaining an efficient and fair duplex \gls*{drl}-\gls*{ris} system for multiple legitimate \gls*{ue} units. We reported experimental work and simulation results, and released our code and experiments to foster further research on this topic.

The results show that fairness-oriented reward functions can significantly improve the equity of resource allocation and mitigate information leakage to eavesdroppers, at the expense of raw system capacity. This trade-off illustrates the difficulty of achieving both high capacity and strong fairness in \gls*{ris}-assisted systems and is consistent with general fair resource allocation theory. 

It is important to acknowledge the inherent sensitivity of deep reinforcement learning to hyperparameters. Fine-tuning such algorithms requires substantial effort, as performance is highly sensitive to hyperparameter choices, network architectures, and training conditions. 

In this study, we therefore restricted our experiments to subcases with dedicated controllers. Based on these limitations and the insights gained from our work, we have identified several promising future research directions, including the development of general controllers that can adapt across diverse network configurations, systematic comparisons with model-based optimization methods, refinements to capture practical \gls*{ris} hardware constraints, and extensions to dynamic scenarios with user and eavesdropper mobility.

\bigskip

\noindent \textbf{Acknowledgments.} The work is supported by the French National Research Agency under the France 2030 label (NF-HiSec ANR-22-PEFT-0009).

\let\clearpage\relax

\vspace{1cm}
\renewcommand{\bibname}{References}
\bibliographystyle{splncs04nat}

\end{document}